# On the new central bank strategy toward monetary and financial instabilities management in finances: Econophysical analysis of nonlinear dynamical financial systems.

Dimitri O. Ledenyov and Viktor O. Ledenyov

*Abstract* – We describe the innovations in finances, introduced over the recent decades, and analyze most of the business and regulatory challenges, faced by the financial industry, because of the present disruptive changes in the global capital markets. We use the integrative thinking approach to formulate the new central bank strategy and propose that the new strategy has to be focused on the constant management of the monetary and financial instabilities, using the knowledge base, accumulated in the field of econophysics. We explain that the econophysics is an emerging scientific discipline that applies the research concepts and methodologies that are originated in the field of physics to understand the nature of problems in the fields of economics and finances, using the nonlinear dynamical analysis, statistical analysis and probabilities theory. We analyze the advances in the field of finances, applying the knowledge base and practical experiences, accumulated in the field of econophysics by the various research groups. We use the integrative thinking approach to formulate the new central bank strategy, which, in our opinion, has to be focused on *the constant management of the monetary and financial instabilities, using the knowledge base in the field of econophysics*. We propose the new theoretical model of economics, which is called the *Nonlinear Dynamic Stochastic General Equilibrium (NDSGE)*, which takes to the account the nonlinearities, appearing during the interaction between the business cycles. We show that the central banks, which will apply the knowledge gained from the econophysical analysis to understand the complex processes in the national financial systems in the time of high volatility in global capital markets, will be able to govern the national financial systems successfully. We show that the national and multinational commercial banks could be best positioned to profit from the innovations in finances in the case, if the new proposed strategy together with the necessary changes of regulatory policies framework would be introduced by the central banks and international regulatory authorities.

The ideas about the realization of prosperous society in the *21* century are widely debated in *Landes (1998)*; *Hara (2012)*. The general opinion is that the well designed **financial system** is a foundation of prosperous society from the political economy point of view in *Hirch (1896)*. The modern global financial system is established due to the development of national and multinational banking in the *UK* and *USA* in *Jones (1993, 2006); Friedman, Schwartz (1971)*; *Rothbard (2002)*. Presently, the established global financial system faces a number of functional challenges and proved to be ineffective, because of some reasons. *Kurose (2003)* writes that the economies of major industrialized countries have been extremely unstable over the recent decades, because the developed economies are exposed to the **uncertainty**. *Itaya (1985)* emphasized that there are the **various types of uncertainty** in a real world: the climate variations, uncertainty about resources available, technological inventions in the future, etc. The uncertainty can be considered as the randomness with the unknown probabilities. It is a well known fact that the growth of **volatility** in the global capital markets is due to the introduction of various advancements and innovations in the field of finances in *Takaishi, Chen, Zheng (2012)* and in *Yoshikawa, Iino, Iyetomi (2012)*. There are many different types of **credit derivatives**, which are considered as the financial innovations and reviewed in *Ledenyov V O, Ledenyov D O (2012)*. The comprehensive analysis of credit derivatives and their applications is conducted in *Hull (2012, 2010, 2005-2006)*. It is a well known fact that the introduced innovations in the field of finances greatly affect the complex financial transactions network, formed by the millions of global firms; that is why the research on the Japanese transaction network, consisting of about 800 thousand firms (nodes) and four million business relations (links) is considered to be very important in *Iino, Iyetomi (2012)*. It is understood that the introduction of financial innovations results in an appearance of the periods of economic growth and recession with the changing volatility dynamics.

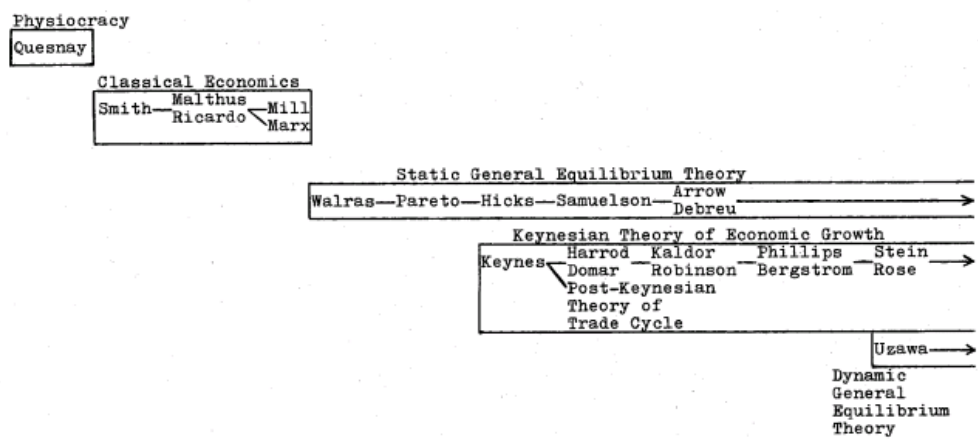

Tab. 1. Paradigms in the History of Economics (after *Kato (1975)*).

These periods of economic growth and recession have the cyclic character. Going from the researched historical facts, it is possible to conclude that the fluctuational nature of economic developments results in an appearance of the various **business cycles** with the certain periodic time-frames. The business cycles may have various periods and amplitudes in *Mosekilde (1996)*. There are the four main business cycles in the developed economies, namely the **Kitchin, Juglar, Kuznets, Kondratiev economic cycles** as explained in *Ledenyov V O, Ledenyov D O (2012)*. *Ikeda, Aoyama, Fujiwara, Iyetomi, Ogimoto, Souma and Yoshikawa* (2012) write that the economists have studied the business cycles phenomenon by the means of mathematical models, including the various kinds of ***linear, nonlinear, and coupled oscillator models***. The Japanese authors state that the business cycle can also be considered as an example of synchronization, which is researched in the ***nonlinear physics*** (by the synchronization, the Japanese researchers mean that there is a constant phase difference between industry sectors) in *Ikeda, Aoyama, Fujiwara, Iyetomi, Ogimoto, Souma and Yoshikawa (2012)*. *Uechi, Akutsu* (2012) explain that a big number of complicated interactive systems, including the financial and economic systems, can be classified as the ***nonlinear dynamical systems***, which play the essential roles in the fields of natural sciences, engineering, economy, and finances. The nonlinearities can originate in the various resonant systems in the fields of electronics, physics and mathematics. For example, the nonlinearities can be observed in the microwave resonators in the field of electronics. The origin of nonlinearities, observed in the microwave resonators in the field of electronics, was comprehensively researched in *Ledenyov D O, Ledenyov V O* (2012). In general, the nonlinear dynamical systems are usually characterized by the self-interactions, self-organizations, spontaneous emergence of order, dissipative structures and nonlinear cooperative phenomena in *Uechi, Akutsu* (2012); Prigogine, Stengers (1984); Kauffman (1993); Mosekilde (1996, 1996-1997); Kuznetsov (2001). The Japanese researchers explain that, in the field of economics, the conserved quantity in the system of business cycle has been researched, regarding this business cycle model as a predator-prey type competitive system, which can be described by ***the predator-prey nonlinear differential equations***, known as the *Lotka-Volterra* (*LV*) equation in *Uechi, Akutsu* (2012). Moreover, the mathematical model for the *Lanchester Strategic Management* is known as a predator-prey type competitive system, which can also be described by the predator-prey nonlinear differential *Lotka-Volterra* (*LV*) equations in *Uechi, Akutsu* (2012). We would like to add that the response function of the nonlinear dynamical system depends on many parameters, which have to be taken to the account during the computer modeling and simulation of the economic and financial systems behaviour. There is a big number of complicated methods from the statistical physics and mathematics, which can be applied to perform the economic and financial analyses with the purpose to predict the market fluctuations, accompanied by the periods of economic recession and stagnation. For example, the Japanese researches apply the mathematical analysis to analyze the complex collective behavior of *stock prices*, which is considered as a possible precursor to the market crash in *Maskawa (2012)*. In some cases, these trends and patterns of stock markets are analyzed, using the different types of random number generators in *Yang, Itoi, Tanaka-Yamawaki (2012)*.

We believe that it is necessary to apply the *integrative thinking* with the aim to develop the new regulatory policies framework and new central bank strategies in the time of high volatility in global capital markets in *Martin (2005-2006, 2008, 2009)*. We have already successfully applied the integrative thinking approach with the purpose to analyze the situation and propose the new innovative solutions to overcome the current financial downturn in North America and in Europe, namely we made the two original proposals on the optimization of functional performances of national and global financial systems in the time of globalization, suggesting the introduction of the *random tax* and *quantum tax* in *Ledenyov V O, Ledenyov D O (2012)*.

In this research paper, we take a few steps forward by proposing the new economic remedies with the purpose to resolve the present financial crisis in the World, going from the knowledge base, gained and accumulated in the field of *econophysics*.

We draw the reader's attention to the well known facts that the central banks in different countries govern the state finances through in *Ledenyov V O, Ledenyov D O (2012)*:
1. The creation of *financial policies* for effective systemic financial regulation;
2. The creation of *monetary base* to enable the various financial operations among the economic agents.

It is a well accepted opinion that there are the two main functional purposes of the central bank such as to establish and maintain in *Ledenyov V O, Ledenyov D O (2012)*:
1. *Monetary stability*, which means stable prices and confidence in the currency;
2. *Financial stability*, which entails for detecting and reducing threats to the financial system as a whole.

The present strategies by the central banks pursue the only goal to reach the *monetary and financial stabilities*. These strategies were implemented through the creation of monetary bases and financial policies to reach the monetary and financial stabilities, and failed to achieve their goals. *We think that the old central banks strategy to reach the monetary and financial stabilities by any possible means have to be disregarded.*

Let us discuss our strategic proposals in finances in details. We state the economic and financial systems can be classified as the nonlinear dynamical fluctuational systems with the dynamic chaos properties, hence they can be predictable to a certain extend. We advocate that the positive and negative feedback loops, which have place in the nonlinear dynamical fluctuational systems, must be taken to the account by the central banks during the consideration of the nonlinear phenomena in the field of finances and economics with the aim to predict the financial and economic systems behaviour. We emphasis that the nonlinear dynamical fluctuational systems are associated with the significant phase delays, and can be characterized by the fluctuations with the big enough amplitudes to generate the nonlinear distortions, which in turn can greatly impact the economic and financial systems, hence they require the detailed consideration by the central banks. ***Therefore, we apply the integrative thinking approach to formulate the new central bank strategy and propose that the new strategy has to be focused on the constant management of the monetary and financial instabilities, using the knowledge base in the field of dynamic chaos in the econophysics.***

Let us make a few additional strategic propositions in the field of economics. The macroeconomics and microeconomic principles and theories, which consider the complex economic problems, have been proposed by many economists and most recently reviewed in Krugman, Wells (2005), Stiglitz (2005), *Ledenyov V O, Ledenyov D O (2012)*. In Japan, *Kato (1975)* considered the various paradigms in the history of economics in Tab. 1 and proposed the ***Dynamic General Equilibrium Theory***, when making his research on the procedure for the dynamization of general equilibrium theory in the terms of calculus of variations. Since that time, the standard theoretical model of modern economics, which is called the ***Dynamic Stochastic General Equilibrium*** (***DSGE***), was developed as noticed in Yong Tao (2012). ***We propose the new theoretical model of economics, which is called the Nonlinear Dynamic Stochastic General Equilibrium model (NDSGE), which takes to the account the appearing nonlinearities during the interaction between the business cycles, using the econophysics methods.*** Let us note that the comprehensive discussion on the *NDSGE* is beyond the scope of this short article, and is conducted in our other research papers. In this article, we can make a comment that the well known research papers and books do not consider the origin of and impact by the nonlinearities in the fields of finances and economics from the econophysics perspective.

We conclude by making the following statements:

1. We have shown through an advanced research that the *econophysics* can help us to better understand the nature of present processes in the fields of finances and economics.

2. We used the integrative thinking approach to formulate the new central bank strategy, which has to be focused on *the constant management of the monetary and financial instabilities, using the knowledge base in the field of econophysics*.

3. We proposed the new theoretical model of economics, which is called the *Nonlinear Dynamic Stochastic General Equilibrium (NDSGE)*, which takes to the account the appearing nonlinearities during the interaction between the business cycles, using the econophysics methods.

Authors are very grateful to the *Yukawa Institute for Theoretical Physics* at *Kyoto University* in Japan for a kind opportunity to get an open access and analyze the research papers, presented at the *YITP* workshop on "*Econophysics 2011 — The Hitchhiker's Guide to the Economy*." We appreciate the *Graduate School of Economics and Business Administration at Hokkaido University* in Japan for giving us a wonderful opportunity to conduct the research on the highly innovative research papers, written by the Japanese scientists in 2012.

The first author appreciates *Prof. Janina E. Mazierska, Electrical and Computer Engineering Department, School of Engineering and Physical Sciences, James Cook University, Australia* for an opportunity to make the advanced innovative research on the nonlinear dynamic resonant systems in the field of microwave superconductivity during more than 12 years.

The second author appreciates *Prof. Roger L. Martin* for the presented opportunity to lean more about the integrative thinking in 2005-2006; and thanks to *Prof. John C. Hull* for the numerous thoughtful discussions on the credit derivatives at the *Rotman School of Management, University of Toronto* in Canada in 2005-2006. *Lionel Barber, Editor-in-Chief, Financial Times* is appreciated for his kind encouragements, including the research opportunity to discuss the complex issues in the field of finances with more than one hundred economists, financiers, professors and journalists in the *FT* in London in the *UK* in recent years.

[*)] This condensed version of our research article is submitted to *The Financial Times, The Bodley Head and The Random House first annual essay competition* in London in the U.K. in 2012.

*E-mail: dimitri.ledenyov@my.jcu.edu.au